\documentclass[twoside]{article}
\usepackage{amssymb}

\newcommand{\prepno}[2]
    {\thispagestyle{empty}
     \noi   \unitlength=1mm
    	\begin{picture}(178,10)
            \put(177,15){\llap{\bf #1}}
            \put(177,10){\llap{\small\rm #2}}
        \end{picture}\bigskip
          }                        

\newcommand{\Title}[1]{\noi {\Large\bf #1} \\}
\newcommand{\Author}[2]{\noi{\large\bf #1}\\[2ex]\noindent{\it #2}\\}
\newcommand{\foom}[1]{\protect\footnotemark[#1]}
\newcommand{\email}[2]{\footnotetext[#1]{e-mail: #2}
		\addtocounter{footnote}{1}}
\newcommand{\Abstract}[1]{\vskip 2mm \begin{center}
        \parbox{16.4cm}{\small\noi #1} \end{center}\medskip}

\makeatletter
\renewcommand{\section}{\@startsection{section}{1}{0pt}%
        {-3.5ex plus -1ex minus -.2ex}{2.3ex plus .2ex}%
        {\large\bf\protect\raggedright}}

\renewcommand{\subsection}{\@startsection{subsection}{2}{0pt}%
        {-3ex plus -1ex minus -.2ex}{1.4ex plus .2ex}%
        {\normalsize\bf\protect\raggedright}}

\renewcommand{\thesubsubsection}%
        {\arabic{section}.\arabic{subsection}.\arabic{subsubsection}.}
\renewcommand{\@oddhead}{\raisebox{0pt}[\headheight][0pt]{%
   \vbox{\hbox to\textwidth{\rightmark \hfil \rm \thepage \strut}\hrule}}}
\renewcommand{\@evenhead}{\raisebox{0pt}[\headheight][0pt]{%
   \vbox{\hbox to\textwidth{\thepage \hfil \leftmark \strut}\hrule}}}
\newcommand{\heads}[2]{\markboth{\protect\small\it #1}{\protect\small\it #2}}
\newcommand{\Acknow}[1]{\subsection*{Acknowledgement} #1}
\makeatother

\def\sect{Sec.\,}
\def\noi{\noindent}
\def\nqq{\hspace{-2em}}
\def\nhq{\hspace{-0.5em}}

\def\cm{\hspace{1cm}}

\def\eq{Eq.\,}
\def\eqs{Eqs.\,}
\def\beq{\begin{equation}}
\def\eeq{\end{equation}}
\def\bear{\begin{eqnarray}}
\def\al{&\nhq}
\def\lal{&&\nqq {}}               
\def\bearr{\bear \lal}
\def\ear{\end{eqnarray}}

\def\tst{\textstyle}
\def\dst{\displaystyle}

\newcommand{\fract}[2]{{\tst\frac{#1}{#2}}}
\def\nn{\nonumber\\ {}}

\def\yy{\\[5pt]}

\def\eql{\al =\al}

\def\eqdef{\stackrel{\rm def}{=}}
\def\e{{\,\rm e}}
\def\eps{\varepsilon}
\def\d{\partial}

\def\sign{\mathop{\rm sign}\nolimits}
\def\diag{\mathop{\rm diag}\nolimits}

\def\const{{\rm const}}

\def\half{{\tst\frac{1}{2}}}

\newcommand{\vars}[1]{\left\{\begin{array}{ll}#1\end{array}\right.}

\def\wide {\vphantom{\dst\int}}

\def\wt{\widetilde}
\def\cR{{\cal R}}
\def\ocR{\overline{\cal R}}
\def\oR{{\overline R}{}}
\def\R{{\mathbb R}}
\def\M{{\mathbb M}}
\def\V{{\mathbb V}}
\def\mN{_M^N}

\def\MN{^{\mu\nu}}
\def\og{{\overline g}}

\def\uc{{\underline c}}
\def\vY{{\vec Y}}
\def\oo{\omega_1}

\def\eom{\eta_\omega}

\def\kappa{\varkappa}

\def\dsJ{\mbox{$ds^2_{\rm J}$}}
\def\cf{c_{\varphi}}
\def\etaF{\eta_{{}_F}}

\heads {K.A. Bronnikov}
{Nonsingular multidimensional cosmologies with Lobachevsky spatial sections}

\begin{document}
\prepno {gr-qc/0410119}
{\it Lecture at Int. School GAS\,04, Kiten, Bulgaria, 10-16 June 2004}

\Title
	{Nonsingular multidimensional cosmologies\yy
	 with Lobachevsky spatial sections }

\Author{Kirill A. Bronnikov\foom 1}
{VNIIMS, 3-1 M. Ulyanovoy, Moscow 119313, Russia, and\\
 Institute of Gravitation and Cosmology, PFUR,
 	6 Miklukho-Maklaya St., Moscow 117198, Russia}
\email 1 {kb@rgs.mccme.ru, \ kb20@yandex.ru}

\Abstract
{Examples of nonsingular cosmological models are presented on the basis    
of exact solutions to multidimensional gravity equations. These examples
involve pure imaginary scalar fields, or, in other terms, ``phantom''
fields with an unusual sign of the kinetic term in the Lagrangian.
We show that, with such fields, hyperbolic nonsingular models with a
cosmological bounce (unlike spherical and spatially flat models) emerge
without special relations among the integration constants, i.e., without
fine tuning. In such models, the extra-dimension scale factors as well
as scalar fields evolve smoothly between different finite asymptotic
values. Examples of theories which create phantom scalar fields are
obtained from string-inspired multidimensional field models and
from theories of gravity in integrable Weyl space-times.

}

 \section{Introduction}

    The recent years have been marked with a lot of violent events in both
    observational and theoretical cosmology. The discovery of an
    accelerated expansion of the Universe \cite{accel} has been one of the
    most important empirical findings. There followed a flood of theoretical
    works trying to interpret and to explain this acceleration, see, e.g.,
    \cite{th-accel}. The majority of such constructions have a common
    feature: they involve various kinds of scalar fields. In many cases
    these are so-called phantom scalars, having a ``wrong'' sign of the
    kinetic term in their Lagrangians.

    It is of interest that such scalar fields are able to suggest a solution
    to one more long-standing problem of theoretical cosmology, namely, the
    initial singularity problem. Inclusion of such fields makes it possible
    to circumvent the well-known singularity theorems and to prevent the
    formation of a cosmological singularity, keeping the curvature at
    sub-Planckian scales.

    In these notes we will discuss in some detail this mechanism of avoiding
    a singularity in isotropic cosmological models at the level of classical
    field theory. Its efficiency is clear from the simplest example of a
    time-dependent scalar field in general-relativistic isotropic cosmology,
    described in \sect 2.  We begin with a discussion of the late-time
    behaviour of various models with accelerated expansion and show that a
    deceleration parameter $q_0< -1$ may be obtained with phantom scalar fields
    without creating a final singularity (``big rip'') related to an
    infinite growth of the scale factor at finite time. It is then shown
    that an initial singularity can also be avoided with the aid of such
    fields: it turns out that a regular minimum of the scale factor is only
    possible in hyperbolic models with a phantom scalar field.

    A similar mechanism works as well in more complex cases to be discussed
    in the further sections. \sect 3 gives examples of multidimensional
    models \cite{J02}, emerging in the field limit of some topical
    string-based unification theories (see \cite{duff,hukhu} and reviews on
    string cosmology \cite{string_cos}). Phantom type fields take
    place in such models with dimensions $D \geq 11$. We will not touch
    upon the whole wealth of exact solutions obtained in such models and
    restrict ourselves to simple models with a single dilatonic field, a
    single axionic type antisymmetric form and two scale factors --- the
    ``external'' one, $a(t)$, and the ``internal'' one, $b(t)$. We shall see
    that nonsingular solutions in which, as $t \to \infty$, the scale factor
    $a(t)$ grows while $b(t)$ and the dilaton $\phi(t)$ tend to finite
    limits, exist in closed and spatially flat cosmologies ($K_0= 0, +1$)
    with some special values of the integration constants only, i.e.,
    require ``fine tuning'', whereas in hyperbolic models with Lobachevsky
    3-geometry ($K_0 = -1$) they emerge generically, without any fine tuning.

    In \sect 4, similar inferences are obtained for cosmological models of a
    certain class of gravitation theories with Weyl non-metricity which are,
    in many observational predictions, equivalent to scalar-tensor
    theories (STT) of gravity and can also contain phantom scalar fields
    \cite{BKM,bm96}. Their main difference from STT is a geometric
    interpretation of the scalar fields which makes their possible
    phantom character more natural than for usual, material scalar fields.

    \sect 5 contains some concluding remarks.

\section{4D scalar field cosmologies at late and early times}

    To illustrate the general properties of scalar field driven cosmologies,
    let us consider the standard 4D Friedmann-Robertson-Walker (FRW) metric
\bear
    ds^2 = -dt^2 + a^2(t) ds_0^2, \cm                  \label{RW}
    ds_0^2 = \frac{dr^2}{1-K_0 r^2} +
    		          r^2 (d\theta^2 + \sin^2\theta d\varphi^2),
\ear
    where $a(t)$ is the scale factor and $ds_0^2$ is the metric of a
    3-space of constant curvature $K_0 = 0,\ \pm 1$. The corresponding
    Einstein-Friedmann equations are
\bear
	\frac{3}{a^2} (a'^2 + K_0)  = \kappa  \rho, \cm	    \label{0,1}
	\frac{1}{a^2} (2aa'' + a'^2 + K_0) = - \kappa p,
\ear
    where $\kappa = 8\pi G$ is the gravitational constant, the prime denotes
    $d/dt$, $\rho = -T^0_0$ is the total density of matter and $p = T^1_1 =
    T^2_2 = T_3^3$ its pressure. The matter can be of any origin,
    but the symmetry of its stress-energy tensor
    $T\mN = \diag (-\rho,\ p,\ p,\ p)$ is determined by the choice of the
    metric (\ref{RW}).

    The most important kinematic observational parameters characterizing the
    Universe evolution are the Hubble parameter $H(t) := a'/a$ and the
    deceleration parameter $q(t) := -aa''/a'^2$.  (The parameter $q$ was
    introduced when it was believed that the expansion of the Universe was
    decelerating; its negative values correspond to an accelerating
    Universe.) The current observational estimates of these two parameters
    considerably vary from one paper to another but can be more or less
    reliably taken as \cite{observ}
\bear                                                      \label{H,q_obs}
     H_0 \approx  0.71 \pm 4\ \frac{\rm km}{\rm s\cdot Mpc}, \cm
     q_0 \approx  -1 \pm 0.4,
\ear
    where the subscript ``0'' refers to the present epoch.

    For $\rho\ne 0$, one can always write $p = w\rho$ where $w$ is in
    general time-dependent. The simplest models are obtained for
    $w = \const$ (the so-called barotropic matter). In this case the
    conservation law $\nabla_\alpha T^\alpha_\mu =0$ leads to the relation
\bear
      \rho = \const\cdot a^{-3(w+1)},                        \label{rho_}
\ear
    giving the well-known laws $\rho\sim a^{-3}$ for dust ($w=0$)
    and $\rho\sim a^{-4}$ for disordered radiation ($w=1/3$). Accelerated
    expansion ($q < 0$, however, requires a negative pressure. It is
    easily seen that, for $w < -1/3$ (which is needed for obtaining $q < 0$)
    and a large scale factor $a$, the term with $K_0$ in (\ref{0,1}) is
    negligible as compared with $\rho$ and $p$, so that {\it the scale factor
    behaviour at late times does not depend on the spatial curvature.\/} The
    $t$ dependence of the scale factor is then described as follows:
\begin{description}
\item[a)]
	if $-1/3 > w > -1$, so that the dominant energy condition holds,
	the expansion may be called power-law inflation:
  \bear                                                       \label{power}
	  a \sim t^{2/[3(w+1)]}, \cm q = -1 + \fract 32 (w+1) >- 1;
  \ear
\item[b)]
	if $w=-1$, which corresponds to a positive cosmological
	constant, $\rho=\const >0$, we obtain exponential inflation:
  \bear
	  a \sim \e^{Ht}, \cm H= \const, \cm q = -1;          \label{expon}
  \ear
\item[c)]
	if $w < -1$, the matter may be called exotic, and we obtain
	hyper-inflation ending with a singularity related to a blowing-up
	scale factor:
  \bear                                                       \label{hyper}
	  a \sim (t_*-t)^{-2/[3|w+1|]}, \cm
	  			q = - 1 - \fract 32 |w+1| < -1,
  \ear
	where $t_*$ is the singular time instant.
\end{description}
    In case c), matter behaves exotically indeed: its density grows as
    the volume grows, and all this ends with a ``big rip'', a singularity at
    finite physical time, where both $a$ and $\rho$ grow infinitely.

    Such a sad future of our Universe may be, however, avoided even if the
    present value of $w$ is smaller than -1 but if $w= p/\rho$ is
    time-dependent. Indeed, suppose that matter (or its dominating part) is
    represented by a scalar field $\phi$ with the Lagrangian
\bear
      L_s = -\half \eps g\MN \phi_{,\mu}\phi_{,\nu} - V(\phi), \label{L_phi}
\ear
    where $\eps = \pm 1$ and $V(\phi)$ is a potential. For $\phi=\phi(t)$ we
    have
\bear
	\rho = \half \eps \phi'^2 + V, \cm                     \label{w_phi}
	p = \half \eps \phi'^2 - V, \cm
	w = p/\rho = -1 + \frac{2\eps \phi'^2}{2V + \eps\phi'^2}.
\ear
    Thus a normal scalar field ($\eps=+1$) with a positive potential $V$
    gives $w > -1$ wheras a phantom scalar field ($\eps=-1$) with $V > 0$
    leads to $w < -1$. However, if at large $t$ the scalar field tends
    rapidly enough to a minimum of the potential, $V_{\min} > 0$, then $\phi'
    \to 0$ and $w\to -1$ as $t\to \infty$; $V_{\min}$ behaves as an
    effective cosmological constant, and accordingly we obtain the de Sitter
    asymptotic (\ref{expon}).

    Let us now discuss a possible nonsingular behaviour of the early
    Universe, such that the scale factor $a(t)$ undergoes a small but
    nonzero regular minimum $a_{\min}$ at some instant $t=t_0$, so that
    $a'(t_0) =0$, $a''(t_0) >0$. \eqs (\ref{0,1}) then give
\beq
    \sign \rho = K_0, \cm  \rho + 3p < 0.                      \label{0,2}
\eeq

    Suppose again that the model dynamics is dominated by a scalar field
    (\ref{L_phi}) and that near $a_{\min}$ the potential is negligible
    compared to the kinetic term, $|V| \ll |\phi'^2|$. Then, for
    $\phi = \phi(t)$, we obtain $p \approx \rho \approx \eps\phi'^2$, and it
    is straightforward to find that the conditions (\ref{0,2}) hold only in
    case $K_0 = -1$, $\eps = -1$. Thus, {\it among massless scalar fields,
    only the ``phantom'' one can lead to a bouncing isotropic cosmology, and
    only in hyperbolic models with Lobachevsky 3-geometry.\/}

    One can note that the kinetic term, whose effective equation of state is
    $p=\rho$, evolves like $a^{-6}$ [see (\ref{rho_})] whereas the potential
    term behaves qualitatively as a cosmological constant, therefore its
    neglection at small $a(t)$ is justified.

\section{A multidimensional $p$-brane model}

\subsection {General features}

    Consider the action of $D$-dimensional gravity interacting with a
    dilatonic scalar field $\Phi$ and antisymmetric forms $F_s$, $F_p$, which
    account for contributions from both the Neveu-Schwarz --- Neveu-Schwarz
    (NS-NS) and Ramond-Ramond (RR) sectors:
\beq  							      \label{SJ}
    S_{\rm J} = \int d^D x \sqrt{g}\biggl\{
	\Phi \biggl[ \cR -\omega \frac{(\d\Phi)^2}{\Phi^2}
	-\sum_{s} \frac{1}{n_s!} F_s^2 \biggr]
	-\sum_{r} \frac{1}{n_r!} F_r^2\biggr\}
\eeq
    where $\cR$ is the scalar curvature, $g = |\det g_{MN}|$,
    $(\d\Phi)^2= g^{MN}\d_M\Phi \d_N \Phi$, $M,N = 0,\ldots, D-1$, $\omega$
    is a (Brans-Dicke type) coupling constant, $n_s$ and $n_r$ are the ranks
    of antisymmetric forms belonging, respectively, to the NS-NS
    and RR sectors of the effective action; for each $n$-form,
    $F_n^2 = F_{n,M_1\ldots M_n} F_n^{M_1\ldots M_n}$.

    The action (\ref{SJ}) is written in the so-called Jordan conformal frame
    where the field $\Phi$ is nonminimally coupled to gravity. This form is
    actually obtained in the weak field limit of many underlying theories
    as the framework describing the motion of fundamental objects,
    therefore we will interpret the metric $g_{MN}$ appearing in (\ref{SJ})
    as the physical metric. Thus, if the fundamental objects are strings,
    one has in any dimension $\omega = - 1$, while in cases where
    such objects are $p$-branes, one finds \cite{duff}
\beq
    \omega = - \frac{(D-1)(p-1) -(p+1)^2}{(D-2)(p-1)- (p+1)^2},  \label{oDd}
\eeq
    where $p$ is the brane dimension and $D$ is the space-time dimension.
    The NS-NS sector of string theory predicts a Kalb-Ramond
    type field with $n_s = 3$; the type IIA superstring effective action
    contains RR terms with $n_r = 2,\ 4$, while type IIB predicts $n_r =
    3,\ 5$. The action (\ref{SJ}) may also represent the bosonic sectors of
    theories like 11-dimensional supergravity (where the dilaton is absent,
    and there is a 4-form gauge field), or 10-dimensional supergravity (there
    is a dilaton and a 3-form gauge field), or 12-dimensional ``field theory
    of F-theory'' \cite{khve}, admitting the bosonic sector of 11-dimensional
    supergravity as a truncation. The model \cite{khve} contains a dilaton
    and two $F$-forms of ranks 4 and 5; it admits electric 2- and 3-branes
    and magnetic 5- and 6-branes.

    The standard transformation
\beq
	g_{MN} = \Phi^{-2/(D-2)} \og_{MN}      		  \label{trans}
\eeq
    leads to a theory reformulated in the Einstein conformal frame, more
    convenient for solving the field equations:
\bear                                                     \label{SE}
    S_{\rm E} = \int d^D x \sqrt{g_{\rm E}}\biggl\{
	\ocR - \eta_{\omega} (\d\varphi)^2
	-\sum_{s} \frac{\eta_s}{n_s!} \e^{2\lambda_{s}\varphi}F_s^2
	-\sum_{r} \frac{\eta_r}{n_r!} \e^{2\lambda_{r}\varphi}F_r^2
	\biggr\}
\ear
    where all quantities are written in terms of the Einstein-frame metric
    $\og_{MN}$; $g_{\rm E} = |\det\og_{MN}|$;
    for the scalar field we have denoted
\beq
	\Phi = \e^{\varphi/\omega_1},\cm                    \label{Phi}
	\omega_1 = \sqrt{\biggl|\omega + \frac{D-1}{D-2}\biggr|};\cm
        \eta_\omega =\sign \biggl(\omega + \frac{D-1}{D-2}\biggr),
\eeq
    while the coupling constants $\lambda_s$ and $\lambda_r$ are
\bear                                                           \label{lamd}
    \lambda_s = \frac{n_s -1}{\omega_1(D-2)}\cm \mbox{(NS-NS sector);}
\cm
    \lambda_r = \frac{2n_r -D}{2\omega_1(D-2)}\cm \mbox{(RR sector).}
\ear

    The sign factor $\eom$ distinguishes ``normal'' theories ($\eom=+1$),
    such that the kinetic term of the $\varphi$ field in (\ref{SE}) has the
    normal sign corresponding to positive energy, from anomalous theories
    where this sign is ``wrong'' ($\eom = -1$). The factor $\eom$ is thus
    quite similar to $\eps$ used in \sect 2. It should be noted that many
    theories with $D > 11$ involve $\eom = -1$. According to (\ref{Phi}),
\beq
    \frac{\eom}{\omega_1^2}
           = (D-2)\biggl[1 - \frac{(D-2)(p-1)}{(p+1)^2}\biggr]. \label{o1}
\eeq
    Evidently, under the condition $(D-2)(p-1) > (p+1)^2$
    we have $\eom =-1$. For $p=2,\ 5$ this happens when $D>11$, and for
    $p= 3,\ 4$ when $D > 10$.

    The following table gives the values of $\omega$ and $\eom/\omega_1^2$
    for some particular space-time and brane dimensions.

\def\z{\phantom{$-$}}
\begin{center}
\begin{tabular}{|c|c|l|l||c|c|l|l|}
\hline
$\wide \  D \ $ & \ $p$ \ & $\omega$ & $\eom/\oo^2$ &
		          \ $D$ \ & \ $p$ \ & $\omega$ & $\eom/\oo^2$ \\
\hline
 any    &  1  &    $-1$    &  $D-2$   & 12  &  2  &   $-2$   &   $-10/9$  \\
 10     &  2  &  \z 0      & \z 8/9   & 12  &  3  &  $-3/2$  &   $-5/2$   \\
 10     &  3  & \z$\infty$ & \z 0     & 12  &  4  &  $-8/5$  &   $-2$     \\
 10     &  4  &  \z 2      & \z 8/25  & 12  &  5  &   $-2$   &   $-10/9$  \\
 10     &  5  &  \z 0      & \z 8/9   & 12  &  6  &   $-6$   &   $-10/49$ \\
 10     &  6  &  $-4/9 $   & \z 72/49 & 12  &  7  & \z 1/2   & \z  5/8    \\
 11     &  2  & \z$\infty$ & \z 0     & 12  &  8  & $-4/11$  & \z 110/81  \\
 11     &  3  &    $-2$    & $-9/8$   & 14  &  2  & $-4/3 $  &$  -4     $ \\
 11     &  4  &    $-5/2$  & $-18/25$ & 14  &  6  & $-16/11$ &$ -132/49 $ \\
 11     &  5  & \z$\infty$ & \z 0     & 26  &  3  & $-17/16$ &$  -48    $ \\
 11     &  6  &  \z 1/4    & \z 36/49 & 26  &  4  & $-50/47$ &$-1128/25 $ \\
\hline
\end{tabular}
\end{center}

    Some comments are in order.
    First, the well-known result $\omega= - 1$ for strings ($p=1$) in any
    dimension is recovered. Second, one obtains $\omega=\infty$ for 2- and
    5-branes in 11 dimensions, which conforms to the absence of a dilaton in
    11D supergravity that predicts such branes.  Third, in 12 dimensions one
    has $\eta_\omega = -1$ for $p<7$, and such a theory \cite{khve}
    does contain a pure imaginary dilaton: the $F$-forms of ranks 4 and 5
    are coupled to a dilaton field $\varphi$ with the coupling constants
    $\lambda_1^2 = -1/10$ and $\lambda_2 = -\lambda_1$, respectively, while
    the product $\lambda\varphi$ is real. As is concluded in Ref.\,\cite
    {khve}, for $D>11$ ``imaginary couplings are exactly what is needed in
    order to make a consistent truncation to the fields of type IIB
    supergravity possible''. In our (equivalent) formulation, $\varphi$ and
    $\lambda$ are real and the unusual nature of the coupling is reflected
    in the sign factor $\eom$. Supersymmetric models with $D=14$ are also
    discussed \cite{bars,gavk}, while $D=26$ is the well-known dimension for
    bosonic strings.

 \subsection{Solutions}

    There is a diversity of exact solutions (discussed, in particular, in
    Refs.\,\cite{k1,k2,ivmel}, see also references therein) for the action
    (\ref{SE}) without $L_m$, in space-times with the metric $\og_{AB}$ of
    the form
\bear
    ds^2_{\rm E}                                               \label{dsE}
	= - \e^{2\alpha(u)} du^2 + \sum_{i=0}^{n} \e^{2\beta^i(u)}ds_i^2
\ear
    where $u$ is a time coordinate and $ds^2_i$ are $u$-independent
    metrics of internal $d_i$-dimensional factor spaces assumed to be
    Ricci-flat for $i=1,\ldots,n$ whereas $ds_0^2$ describes the ``external''
    (observed) space of constant curvature $K_0 = 0, \pm 1$, corresponding
    to the three types of isotropic spaces.

    We will be only interested here in cosmological solutions for a very
    simple special case: a single antisymmetric form $F_{[d_0]}$ from the
    NS-NS or RR sector, having a single (up to permutations) nontrivial
    component $F_{1...d_0}$ where the indices refer to the external
    space $\M_0$ with the metric $ds_0^2$, a single internal space $\M_1$
    with the metric $ds_1^2$, so that in (\ref{dsE}) $i=0,1$, and $\varphi =
    \varphi(u)$.  Then the field equations are easily integrated.

    Let $u$ be a harmonic time coordinate for the metric (\ref{dsE}), so
    that the coordinate condition is $\alpha = d_0 \beta^0 + d_1\beta^1$.

    The $F$-form is magnetic-type; the Maxwell-like equations due to
    (\ref{SE}) are satisfied trivially while the Bianchi identity
    $dF=0$ implies
\beq
	F_{1...d_0} = Q \sqrt{g_0}, \cm Q = \const,           \label{Q}
\eeq
    where $g_0$ is the metric determinant corresponding to $ds^2_0$
    and $Q$ is a charge, to be called the {\it axionic charge\/} since the
    only nonzero component of $F$ can be represented in terms of a
    pseudoscalar axion field in $d_0+1$ dimensions. The remaining unknowns
    are $\beta^0$, $\beta^1$ and $\varphi$.

    In the Einstein equations $\ocR\mN - \half \delta\mN \ocR = T\mN$,
    written for the Einstein-frame metric (\ref{dsE}),
    the stress-energy tensor $T_M^N$ has the property $T^u_u + T_z^z = 0$
    (where $z$ belongs to $\M_0$), and the corresponding Einstein equation
    has the Liouville form $\ddot\alpha - \ddot \beta^0 + K_0 (d_0-1)^2
    \e^{2\alpha-2\beta^0}$, whence
\bearr
    \frac{1}{d_0-1}\e^{\beta_0-\alpha}                          \label{S}
    = S(-K_0,k,u) \eqdef \vars{\e^{ku},   & K_0 =0,\quad k\in\R;\\
			  k^{-1}\cosh ku, & K_0=1,\quad k>0;\\
			  k^{-1}\sinh ku, & K_0 =-1, \quad k> 0;\\
			  u,              & K_0 = -1,\quad k=0;\\
			  k^{-1}\sin ku,  & K_0 =-1, \quad k < 0, }
\ear
    where $k$ is an integration constant and one more constant is suppressed
    by a proper choice of the origin of $u$. \eq(\ref{S}) can be used to
    express $\beta^0$ in terms of $\beta \equiv \beta^1$.

    It is helpful to consider the remaining unknowns
    as a vector $x^A = (\beta^1,\ \varphi)$
    in the 2-dimensional target space $\V$ with the metric
\beq
    (G_{AB}) = \pmatrix{        d\,d_1 & 0    \cr    	      \label{GAB}
	        	 	     0 & \eom \cr}, \cm
    (G^{AB}) = \pmatrix{    1/(d\,d_1) & 0    \cr
	        	 	     0 & \eom \cr},
		\cm   d \eqdef \frac{D-2}{d_0-1}.
\eeq
    The equations of motion then take the form
\bearr                                                       \label{eq-x}
    \ddot x{}^A = - \etaF Q^2 Y^A \e^{2y}
\\  \lal                                                         \label{int}
    G_{AB}\dot x^A \dot x^B + \etaF Q^2 \e^{2y} = \frac{d_0}{d_0-1}K,
    \cm
	 K = \vars {k^2\sign k, & K_0 = -1,\\
		    k^2,        & K_0 = 0, +1.}
\ear
    with the function $y(u) = d_1 \beta^1 + \lambda\varphi$, representable
    as a scalar product of $x^A$ and the constant vector $\vY$ in $\V$:
\beq
    y(u) = Y_A x^A, \cm  Y_A = (d_1,\ \lambda),  \cm
    		         Y^A = (1/d,\ \eom\lambda).       \label{YA}
\eeq
    \eq (\ref{int}) is a first integral of (\ref{eq-x}) that
    follows from the ${u \choose u}$ component of the Einstein equations.

    The simplest solution corresponds to $Q=0$ (scalar vacuum):
\beq
    \beta^1 = c^1 u + \uc^1, \cm                            \label{vac}
    \varphi = \cf u + \uc_\varphi,
\eeq
    where $c^1,\ \uc^1,\ \cf$ and $\uc_\varphi$ are integration constants.
    Due to (\ref{int}), the constants $c^A = (c^1,\ \cf)$ are related by
\beq                                                     \label{int-vac}
    c_A c^A = dd_1 (c^1)^2 + \eom \cf^2 = \frac{d_0}{d_0-1}K.
\eeq

    If $Q\neq 0$, \eqs (\ref{eq-x}) combine to yield an easily solvable
    (Liouville) equation for $y(u)$:
\beq
    \ddot y + \etaF Q^2 Y^2 \e^{2y}=0,                       \label{eq-y}
	\cm    Y^2 = Y_A Y^A = d_1/d + \eom \lambda^2.
\eeq
    This is a special integrable case of the equations considered, e.g., in
    Refs.\,\cite{k1,k2,ivmel}.
    Assuming\footnote
{Even for $\eom=-1$ one has $Y^2 >0$ for fields from the NS-NS sector in any
 dimension and for fields from the RR sector if $D < 17$.}
    {}$Y^2 > 0$, \eq (\ref{eq-y}) gives
\beq
    \e^{-y(u)} = h^{-1}|Q|Y \cosh [h(u+u_1)]                \label{y}
\eeq
    where $Y = |Y^2|^{1/2}$, $h >0 $ and $u_1$ are integration constants. The
    unknowns $x^A$ are expressed in terms of $y$ as follows:
\beq
    x^A = \frac{Y^A}{Y^2} y(u) + c^A u + \uc^A               \label{xA}
\eeq
    where the constants $c^A=(c^1,\ c_\varphi$ and $\uc^A =(\uc^1,\
    \uc_\varphi)$ satisfy the orthogonality relations
\beq
    c^A Y_A =0,\cm      \uc^A Y_A =0.                        \label{c-ort}
\eeq
    Finally, the constraint (\ref{int}) leads to one more relation
    among the constants:
\beq
     \frac{h^2}{Y^2} + c_A c^A = \frac{d_0}{d_0-1}K.         \label{int2}
\eeq

\subsection {Analysis of cosmological models}

    In what follows, we put $d_0=3$, so that $d_1 = D-4$,
    and identify, term by term, the Jordan-frame metric \dsJ\ obtained in
    the above notations (\ref{trans}), (\ref{dsE}),
\beq
    \dsJ =                                                    \label{dsJ1}
     	  \exp\biggl[-\frac{2\varphi}{\omega_1(D-2)}\biggr]
    \biggl\{
 	  \frac{\e^{-d_1\beta^1}}{2S(-K_0,k,u)}
          \biggl[ \frac{-du^2}{4S^2(-K_0,k,u)} + ds_0^2\biggr]
 				+ \e^{2\beta^1} ds_1^2
 	          				\biggr\},
\eeq
    where the function $S(.,.,.)$ is defined in (\ref{S}), with the familiar
    form of the metric
\beq
    \dsJ = -dt^2 + a^2(t) ds_0^2 + b^2(t) ds_1^2,             \label{dsJ2}
\eeq
    so that $a(t)$ and $b(t)$ are the external and internal scale factors
    and $t$ is the cosmic time.

    To select nonsingular models, let us use the Kretschmann scalar ${\cal K}
    = R_{MNPQ}R^{MNPQ}$, which is in our case a sum (with positive
    coefficients) of squares of all Riemann tensor components
    $R_{MN}{}^{PQ}$. Thus as long as $\cal K$ is finite, all algebraic
    curvature invariants of this metric are finite as well. For the metric
    (\ref{dsJ2}) with $d_0=3$ one has (the primes denote $d/dt$):
\beq
    {\cal K} = 4\biggl[ 3\biggl(\frac{a''}{a}\biggr)^2      \label{Krch}
		    +d_1 \biggl(\frac{b''}{b}\biggr)^2
		+ 3 d_1 \biggl(\frac{a'b'}{ab}\biggr)^2\biggr]
	    +2\biggl[
	          6\biggl(\frac{K_0+{a'}^2}{a^2}\biggr)^2
		+ d_1(d_1-1)\,\frac{{b'}^4}{b^4}\biggr].
\eeq

    By (\ref{Krch}), ${\cal K} \to\infty$
    and hence the space-time is singular when $a\to 0$, $a\to\infty$,
    $b\to 0$ or $b\to\infty$ at finite proper time $t$.
    Accordingly, our interest will be in the asymptotic behaviour of the
    solutions at both ends of the range $\R_u = (u_{\min},\ u_{\max})$ of
    the time coordinate $u$, defined as the range where both $a^2$ and
    $b^2$ in (\ref{dsJ2}) are regular and positive. (Note that, as
    $t\to \pm\infty$, a singularity does not occur when $b(t)\to 0$, or
    $a\to 0$ in case $K_0=0$.) At any $u\in \R_u$ all the relevant functions
    are manifestly finite and analytical. The boundary values $u_{\max}$
    and $u_{\min}$ may be finite or infinite; a finite value of $u_{\max}$
    or $u_{\min}$ coincides with a zero of the function (\ref{S}).

    Among regular solutions, of utmost interest are those in which
    $a(t)$ grows while $b(t)$ tends to a finite constant value as
    $t\to\infty$. Any asymptotic may on equal grounds refer to the
    evolution beginning or end due to the time-reversal invariance of the
    field equations. We will for certainty speak of expansion or inflation,
    bearing in mind that the same asymptotic may mean contraction
    (deflation).

    Let us now enumerate the possible kinds of asymptotics.

\medskip\noi
    {\bf Type I:} $\ u\to \pm \infty$,
\beq
    dt^2 \sim \e^{(A-2k)|u|}du^2, \qquad
     a^2 \sim \e^{A|u|}, \qquad                         \label{type1}
     b^2 \sim \e^{B|u|},
\eeq
    with $k>0$ and the constants $A$ an $B$, depending on the parameters of
    the solution. An asymptotic of interest for $a(t)$ takes place if
    $A\geq 2k > 0$:
\begin{description}\itemsep -2pt
\item[(i)]
	$A>2k$: \ \ $t\to\infty$, \ \
	            $a\sim t^{A/(A-2k)}$\ (power-law inflation);
\item[(ii)]
	$A=2k$: \ \ $t\sim |u|\to \infty$,\ \
	            $a\sim \e^{kt}$\ (exponential inflation).
\end{description}
    A reformulation for $k<0$ is evident. The scale factor $b(t)$
    tends to a finite limit if $B=0$, i.e., under a special condition on the
    model parameters (fine tuning).

\medskip\noi
    {\bf Type Ia:} a modification of type I when $k=0$, so that at
    $u\to\infty$
\beq
     dt^2 \sim u^{-3}\e^{Au}du^2,\cm
     a^2 \sim u^{-1}\e^{Au}du^2,\cm  b^2 \sim \e^{Bu}       \label{type1a}
\eeq
    If $A > 0$, we have, as desired, $t\to\infty$ and $a\to\infty$; the
    expansion may be called ``slow inflation" since it is only slightly
    quicker than linear: the derivative $da/dt \sim u$, which behaves
    somewhat like $\ln t$. If $A \leq 0$, then $a\to 0$ at finite $t$
    (singularity). As for $b(t)$, one may repeat what was said in case I.

\medskip\noi
    {\bf Type II:} $u\to 0$, where the function (\ref{S}) tends to zero,
    so that $S(-K_0,k,u) \sim u$, while other quantities involved are
    finite. In this case
\beq
    dt^2 \sim 1/u^3, \cm a^2 \sim 1/u, \cm b^2 \to \const>0. \label{type2}
\eeq
    According to (\ref{type2}), $t\to \pm \infty$, $a(t) \sim |t|$ (linear
    expansion or contraction), whereas both $b(t)$ and $\varphi(t)$ tend
    to finite limits since they do not depend on $S(-K_0, k, u)$.

    The dilaton $\varphi$ in all cases behaves like $\ln b(t)$, but, in
    general, with another constant $B$ in each particular solution.

    This exhausts the possible kinds of asymptotics for $Y^2 > 0$. Solutions
    with $Y^2 \leq 0$, which can emerge when $\eom = -1$, may have other
    asymptotics, but they are of lesser interest.

\subsubsection*{Scalar-vacuum cosmologies}

    The scalar-vacuum models (\ref{dsJ1}), (\ref{vac}) depend on two input
    constants, $D$ (or $d_1=D-4$, or $d=(D-2)/2$) and $\omega$ (or
    $\omega_1$) and three integration constants $k,\ c^1,\ \cf$ related by
    (\ref{int-vac}); two more constants, $\uc^1$ and $\uc_{\varphi}$, only
    shift the scales in $\M_1$ and along the $\varphi$ axis and do not
    affect the qualitative behaviour of the models.

\medskip\noi
{\bf  Closed models, $K_0=+1$}. In this case in (\ref{dsJ1}) $S = k/\cosh
    ku,\ k>0$, hence the solution has two type I asymptotics at $u\to \pm
    \infty$, with $k>0$ and the following constants $A = A_{\pm}$:
\beq                                                            \label{Av+}
    A_{\pm} = -k
    		\mp \biggl[d_1 c^1 + \frac{\cf}{d\omega_1}\biggr],
\eeq
    so that at least at one of the asymptotics $A<0$ whence $a\to 0$ at
    finite $t$, a singularity. The behaviour of $b(t)$ is also singular.

\medskip\noi
{\bf Spatially flat models, $K_0=0$.} One has simply
\bear
     a^2(t) = \e^{Au},                                     \label{Av0}
     			\cm dt \sim \e^{(A-2k)u/2}du,
\ear
    where $A = - \cf/(d\omega_1) - d_1 c^1-k$,  $k\in\R$,
    and again $b^2(t) =\e^{Bu},\ B=\const$. Thus each of the scale factors
    is either constant, or evolves between zero and infinity, and $a=0$
    occurs at finite $t$.

\medskip\noi
{\bf Hyperbolic models, $K_0=-1$.} If $k>0$ [note that, when $\eom=1$, there
    is necessarily $k > 0$ due to (\ref{int-vac})], one has in (\ref{dsJ1})
    $S=k^{-1}\sinh ku$.  Hence the model evolves between a type I asymptotic
    at $u\to\infty$, with $A$ coinciding with $A_+$ in \eq (\ref{Av+}), and
    type II at $u=0$. Since type II is regular, a necessary condition for
    having a nonsingular model is $A\geq 2k$.

    To find out if and when it happens for $\eom = +1$, it is convenient
    to introduce, instead of the two constants $c^1$ and $\cf$ connected by
    (\ref{int-vac}), an ``angle'' $\theta$ such that
\beq
    -c^1 = \sqrt{\frac{3}{2dd_1}} k\,\cos\theta,\cm          \label{theta}
    -\cf = \sqrt{\frac{3}{2}} k\, \sin\theta.
\eeq
    The condition $A\geq 2k$ will be realized for a certain choice of the
    integration constants if $A_+$ given by (\ref{Av+}) has, as a function
    of $\theta$, a maximum no smaller than $2k$. An inspection shows that
    it happens if
\beq
    \omega_1^2 \leq 1/[d(6d - d_1)] = 1/[(D-1)(D-2)].        \label{omax}
\eeq
    This is the only example of a nonsingular (bouncing) vacuum model with
    $\eom=+1$.

    In case $k>0,\ \eom = -1$, a choice of $\cf$ and $c^1$ subject to
    (\ref{int-vac}) such that $A>2k$ is easily made for any $\omega_1$.

    For $\eom = -1,\ k=0$, the model evolves between type Ia and II
    asymptotics, where at the Ia end ($u\to\infty$)
\beq
	A = -d_1 c^1 - \cf/(d\omega_1),      \cm              \label{ABv-}
	B = 2 c^1 - \cf/(d\omega_1).
\eeq
    The necessary condition for regularity, $A > 0$, is satisfied for
    proper $c_1$ and $\cf$ which can be chosen without problems.

    In case $\eom = -1,\ k < 0$, the function $(\ref{S})$ is simply
    $|k|^{-1} \sin |k|u$, and the model has two type II asymptotics at
    adjacent zeros of $S$, say, $u=0$ and $u=\pi/|k|$. This model is
    automatically nonsingular for any further choice of integration
    constants.

    We conclude that among vacuum models only some hyperbolic ones can
    be nonsingular. For $\eom = +1$ in such a case $a(t)$ evolves from
    linear decrease to inflation, or from deflation to linear growth. Only
    in the latter case both $b(t)$ and $\varphi$ tend to finite limits as
    $t\to\infty$ without any fine tuning.

    For $\eom=-1$ there is a model interpolating between two
    asymptotics of the latter kind. Thus, as $t$ changes from $-\infty$ to
    $+\infty$, $a(t)$ bounces from linear decrease to linear increase
    (generically with a different slope) whereas $b(t)$ and $\varphi(t)$
    smoothly change from one finite value to another. The latter model
    exists for generic values of the integration constants.

\subsubsection*{Cosmologies with an axionic charge}

    The solution contains, in addition to the input parameters $D$,
    $\omega$ and $\lambda$, three independent essential integration
    constants: the ``scale parameter'' $k$, the charge $Q$ and also $h$ and
    $\cf$ connected by (\ref{int2}); the constant $c^1$ is excluded by
    the first relation (\ref{c-ort})
\beq
    d_1 c^1 + \lambda \cf =0                                   \label{c1}
\eeq
    so that the quantity $c^A c_A$, appearing in (\ref{int2}), is expressed
    as $c^A c_A = \eom (d/d_1) \cf^2 Y^2$. The fourth constant, the
    ``shift parameter'' $u_1$, as well as $\uc^1$ and $\uc_\varphi$,
    connected by (\ref{c-ort}), are qualitatively inessential.

    Let us begin with ``normal'' models, $\eom = +1$.
    The solution (\ref{xA}) has the form
\bear
    \beta^1 (u) = \frac{1}{dY^2} y(u) + c^1 u +\uc^1,          \label{b1,f1}
\cm
    \varphi(u) = \frac{\lambda}{Y^2}y(u) + \cf u + \uc_\varphi.
\ear
    where the function $y(u)$ is fiven by (\ref{y}). The condition
    (\ref{int2}) leads to $k > 0$ and strongly restricts the possible model
    behaviour. Thus, it can be shown \cite{J02} that for all
    $K_0$ one of the asymptotics belongs to type I with the constants
\bear                                                         \label{AB}
    A \eql -k + \frac{h}{dY^2}\biggl(d_1 + \frac{\lambda\eom}{\oo}\biggr)
	      + \cf \biggl(\lambda - \frac{1}{d\oo}\biggr),
\nn
    B \eql \frac{h}{dY^2}\biggl(-2 + \frac{\lambda\eom}{\oo}\biggr)
		- \cf \biggl(\frac{2\lambda}{d_1} + \frac{1}{d\oo}\biggr).
\ear
    The necessary condition for regularity $A\geq 2k$ may be fulfilled
    for small $\omega_1$, satisfying the condition (\ref{omax}), just as in
    the vacuum case. The constant $B$ may then have any sign, and only a
    special choice of the ratio $\cf/h$ (fine tuning) can lead to $B=0$,
    providing a finite limit of $b(t)$.

    The second asymptotic depends on $K_0$. For closed and flat models it
    is again type I, and a nonsingular behaviour is again acheived by fine
    tuning. For $K_0=-1$, the second asymptotic belongs to type II and is
    always regular. Thus particular models with an axionic charge may be
    even regular for $\eom = 1$) and, in addition, may be inflationary as
    $t\to\infty$.

    The ``anomalous'' models with $\eom =-1$, just as in the vacuum case,
    are more diverse due to arbitrariness in the sign of $k$. For models
    $K_0= 0,\ +1$, as well as for hyperbolic ones with $k \geq 0$, the
    restriction (\ref{omax}) is no more valid, but the type I and Ia
    asymptotics are again only nonsingular for special values of the
    parameters.

    Lastly, the models $K_0 = -1,\ \eom =-1,\ k < 0$, as their vacuum
    counterparts, interpolate between two type II asymptotics and have the
    same qualitative features.

\section {Vacuum multidimensional models with integrable Weyl geometry}

    Let us discuss multidimensional cosmological models assuming that
    space-time possesses $D$-dimensional Weyl geometry characterized by the
    metric $g_{MN}$ and the connection
\beq
     \Gamma^A_{BC}= \wt{\Gamma}{}^A_{BC}
     -\half(\sigma_B\delta^A_C + \sigma_C\delta^A_B-g_{BC}\sigma^A)
								\label{Conn}
\eeq
    where $\wt{\Gamma}{}^A_{BC}$ are the Christoffel symbols for the
    metric $g_{AB}$, $\sigma$ is a scalar field and $\sigma_A = \d_A\sigma$.
    The gravitational field is described by the tensor $g_{AB}$ and the
    scalar $\sigma$, as in scalar-tensor theories (STT). Just as in STT,
    the gravitational Lagrangian may contain different invariant
    combinations of $g_{AB}$ and $\sigma$. Restricting ourselves to
    Lagrangians which are linear in the curvature and quadratic in
    $\sigma_A$, we can write:
\beq
	L = f(\sigma)\cR - h(\sigma)\sigma^A\sigma_A -
    		2\Lambda(\sigma) + L_m \label{Lagr}
\eeq
    where $\cR$ is the Weyl scalar curvature, obtained from the connection
    (\ref{Conn}), $f,\ h$ and $\Lambda$ are arbitrary functions, and $L_m$
    is the non-gravitational matter Lagrangian.

    The field equation are simplified if one expresses the Weyl curvature
    $\cR$ in terms of the Riemannian curvature $R$, corresponding to the
    metric $g_{AB}$:
\beq
      \cR = R - (D-1)\Box \sigma
                -\frac{1}{4}(D-1)(D-2)\sigma^A\sigma_A    \label{Rtransit}
\eeq
    ($R$ and $\Box$ are formed from the Riemannian connection
    $\wt{\Gamma}{}^A_{BC}$) and passes to the Einstein conformal picture
    with the aid of the transformation $g_{MN}= f^{-2/(D{-}2)}\og_{MN}$.
    Omitting a total divergence, we arrive at the following form of the
    Lagrangian:
\beq
     \overline{L} = \oR -
        F(\sigma)\og^{AB}\sigma_A\sigma_B
        + f^{-D/(D{-}2)}[- 2\Lambda(\sigma)+L_m]          \label{Lagr1}
\eeq
    where $\oR$ is the Riemannian scalar curvature for the metric
    $\og_{AB}$, $A_\sigma\equiv dA/d\sigma$ and
\beq
        F(\sigma) = \frac{1}{f^2}\biggl [fh -
                (D{-}1)f\biggl(f_\sigma+\frac{D{-}2}{4}\biggr)+
                \frac{D{-}1}{D{-}2}f_\sigma^2 \biggr].    \label{DefF}
\eeq

    Consider  vacuum ($L_m = 0$) cosmological models with the metric
    (\ref{dsE}), assuming $\Lambda\equiv 0$. One can easily find that the
    substitution $\sigma \mapsto \varphi$, such that
    $d\varphi/d\sigma = \sqrt{|F(\sigma)|}$, leads the action with the
    Lagrangian (\ref{Lagr1}) to a form coinciding with (\ref{SE}) without
    an $F$-form and with the sign factor $\eom$ replaced with
    $\sign F(\sigma)$. Therefore, in the Einstein picture, the vacuum
    cosmologies of the gravitation theory with Weyl integrable space-time
    are entirely identical to the scalar-vacuum models from \sect 3
    (both types of models are equivalent to those of multidimensional
    general relativity with a massless, minimally coupled scalar field).

    A difference can appear after a transition to Jordan's picture since
    the conformal factors $\Phi^{-2/(D-2)}$ in \sect 3 and $f^{-2/(D-2)}$ in
    the present section are, in general, different. However, if one
    supposes that the function $f(\sigma)$ is finite and smooth in the whole
    essential range of $\sigma$, then the qualitative properties of the
    models (as regards their regular or singular behaviour) coincide in the
    Einstein and Jordan pictures. Accordingly, in both pictures, the
    conclusion that hyperbolic models in the presence of a phantom scalar
    field contain a class of nonsingular bouncing models, preserves its
    generality. Such models do not require any fine tuning, and in all of
    them both the scalar field and the extra-dimension scale factors change
    in finite limits.

    Stability of the qualitative features of these models with respect to
    addition of other kinds of matter has been confirmed by a numerical
    study of Weyl cosmologies with one more scalar field representing
    ordinary matter \cite{BKM}.

\section {Concluding remarks}

    Phantom scalar fields are rather widely discussed as one of the dark
    energy candidates, able to explain the present accelerated expansion
    of the Universe, see \sect 2 and, in more detail, e.g., \cite{ph-dark}.
    Such fields, if any, may also dominate in the early Universe, at small
    values of the scale factor $a(t)$, above all, due to maximum stiffness
    ($\rho = p$) of the equation of state of a massless scalar field, so
    that their energy density $\rho$ grows with falling $a(t)$ more rapidly
    than for other kinds of matter: e.g., in 4D FRW models one has $\rho
    \sim a^{-6}$. For the same reason, massive scalar fields, or fields with
    potentials, should actually behave as massless ones at small $a(t)$.
    Therefore the above conclusion that an initial cosmological singularity
    may be avoided without fine tuning in hyperbolic models ($K_0=-1$) due
    to phantom scalar fields seems to be rather general.

    This conclusion proves to be even more important in multidimensional
    cosmologiies. If extra dimensions, being an inevitable ingredient in
    modern unification theories, are considered dynamically, the singularity
    problem becomes even more involved since, in addition to the usual
    cosmological scale factor, the extra dimensions can collapse or blow up,
    leading to a curvature singularity. However, in open models of the type
    discussed here, the external scale factor $a(t)$ dynamically differs
    from other variables. Formally, this circumstance was described here
    by the last line in \eq (\ref{S}): $a^2(t) \sim 1/S(-K_0,k,u)$ tends to
    infinity at two finite values of the harmonic coordinate $u$, which
    correspond to infinite physical time, whereas all other variables, such as
    scalar fields and internal scale factors, remain finite at these values
    of $u$.

    The above simple models certainly do not pretend to describe the whole
    evolution, but only try to guess the qualitative features of the
    bouncing process near the maximum density state. The well-known features
    of standard cosmology: inflation, nucleosynthesis, particle creation etc.
    may follow at later times, but the mechanism described here seems to
    automatically provide stable compactification of the extra-dimension
    scale factors (if any) and constant values of scalars which may be
    related to coupling constants in unification theories, such as the
    dilaton in string theory. Cosmological and astrophysical problems
    related to stable compactification are discussed in Ref.\,\cite{zhuk03},
    see also references therein.

    I would also like to mention that such an exotic matter as a phantom
    scalar field, violating all standard energy conditions, in case it
    is concentrated in comparatively small regions of space, is precisely
    what is needed to create wormholes --- and hence maybe also time
    machines \cite{thorne}. There are many exact wormhole solutions
    involving phantom scalars (those in Refs.\,\cite{br73,h_ellis73} are
    probably the earliest). If such fields do exist in nature, whatever
    be their origin, they are, in principle, a ready construction material
    for wormholes.

    Observations are known to yield the total cosmological density factor
    $\Omega_0$ smaller or close to unity; meanwhile, the presently popular
    spatially flat cosmologies, most convenient for various calculations,
    require the precise equalty $\Omega_0=1$, actually a sort of fine tuning.
    It is much more probable that the real Universe at least slightly
    violates this special requirement, leading to $K_0=-1$ if $\Omega_0<1$.

\Acknow
   {The author thanks Julio Fabris, Vitaly Melnikov and Plamen
   Fiziev for numerous helpful discussions and the organizers of the
   International School GAS04 for kind hospitality. The work was supported 
   in part by ISTC Project 1655.}


\end{document}